\documentclass[10pt,journal,twocolumn]{IEEEtran}
\IEEEoverridecommandlockouts
\usepackage{cite}
\usepackage{amsmath,amssymb,amsfonts,booktabs}
\usepackage{mathrsfs}
\usepackage{graphicx}
\usepackage{textcomp}
\usepackage{xcolor}
\usepackage{enumerate}
\usepackage{epstopdf}
\usepackage{subcaption}
\usepackage{amsthm}

\newtheorem{proposition}{Proposition}

\usepackage{color}
\usepackage{multicol}
\usepackage{verbatim}
\newcounter{mytempeqncnt}

\usepackage{algorithm}
\usepackage{algorithmic}
\allowdisplaybreaks[4]

\def\BibTeX{{\rm B\kern-.05em{\sc i\kern-.025em b}\kern-.08em
		T\kern-.1667em\lower.7ex\hbox{E}\kern-.125emX}}
\begin{document}

\title{
Movable-Antenna Array Empowered ISAC Systems for Low-Altitude Economy
}
\author{
\IEEEauthorblockN{ {\small
Ziming Kuang\IEEEauthorrefmark{1}\IEEEauthorrefmark{2},
Wenchao Liu\IEEEauthorrefmark{1}\IEEEauthorrefmark{3},
Chunjie Wang\IEEEauthorrefmark{1}\IEEEauthorrefmark{2},
Zhenzhen Jin\IEEEauthorrefmark{3},
Jinke Ren\IEEEauthorrefmark{4},
Xuhui Zhang\IEEEauthorrefmark{4}\IEEEauthorrefmark{1},
and Yanyan Shen\IEEEauthorrefmark{1}}\IEEEauthorrefmark{5}\\}
\vspace{3.5pt}
\IEEEauthorblockA{ \footnotesize
\IEEEauthorrefmark{1}Shenzhen Institute of Advanced Technology, Chinese Academy of Sciences, Shenzhen, China\\
\IEEEauthorrefmark{2}University of Chinese Academy of Sciences, Beijing, China\\
\IEEEauthorrefmark{3}Southern University of Science and Technology, Shenzhen, China\\
\IEEEauthorrefmark{4}The Shenzhen Future Network of Intelligence Institute and School of Science and Engineering, The Chinese University of Hong Kong, Shenzhen, China\\
\IEEEauthorrefmark{5}Shenzhen University of Advanced Technology, Shenzhen, China\\
\vspace{-0.75em}
}
\thanks{This work was supported in part by the Shenzhen Science and Technology Program under Grant JCYJ20220818101607015, the Foundation of Key Laboratory of System Control and Information Processing, Ministry of Education, Shanghai, China under Grant Scip20240114, and the National Natural Science Foundation of China under Grant 61503368.
(\emph{Z. Kuang and W. Liu contributed eqaully to this work.}) (\emph{Corresponding authors: Y. Shen} (e-mail: yy.shen@siat.ac.cn) \emph{and X. Zhang} (e-mail: xu.hui.zhang@foxmail.com))
}
\vspace{-18.5pt}
}
\maketitle

\begin{abstract}
This paper investigates a movable-antenna (MA) array empowered integrated sensing and communications (ISAC) over low-altitude platform (LAP) system to support low-altitude economy (LAE) applications.
In the considered system, an unmanned aerial vehicle (UAV) is dispatched to hover in the air, working as the UAV-enabled LAP (ULAP) to provide information transmission and sensing simultaneously for LAE applications.
To improve the throughput capacity and meet the requirement of the sensing beampattern threshold, we formulate a data rate maximization problem by jointly optimizing
the transmit information and sensing beamforming, and the antenna positions of the MA array.
Since the data rate maximization problem is non-convex with highly coupled variables,
we propose an efficient alternation optimization based algorithm, which iteratively optimizes parts of the variables while fixing the others.
Numerical results show the superiority of the proposed MA array-based scheme in terms of the achievable data rate and beamforming gain compared with two benchmark schemes.
\end{abstract}


\section{Introduction}
As an emerging economic paradigm that deeply integrates next-generation communications and Internet of things, low-altitude economy (LAE) has attracted widespread attention from academia and industry \cite{cheng2024networked}.
LAE is composed of low-altitude flight and deployment activities, such as electric vertical take-off and landing, and low-altitude platforms (LAPs).
Although LAE can be applied in fields such as transportation, environmental monitoring, and agriculture, ensuring efficient and robustness communications among low altitude deployment devices and ground users still remains a challenge.
To address this issue, unmanned aerial vehicles (UAVs)-enabled LAP (ULAP) is proposed, benefited by their advantages of flexibility in deployment, stability in line-of-sight (LoS) links, and controllable mobility \cite{8918497}.

However, traditional ULAPs can only provide simple transmissions, which are difficult to meet the diverse requirements of ground users, including the ultra-reliable and low-latency wireless connectivity, and the highly accuracy and environmental-robustness sensing capability \cite{9737357}.
Fortunately, integrated sensing and communications (ISAC) is proposed, which fully combines dedicated sensing and communication functionalities to efficiently utilize 
wireless and hardware resources \cite{9606831}.
Since commercial UAVs are equipped with a series of sensors, enabling ISAC on ULAPs becomes a feasible solution.
Besides, differ from the sensor-based sensing, radio-based sensing is also utilized into ULAPs, where the detection and estimation of the targets-of-interest are performed on the radio signals echoed and/or scattered by the targets \cite{9456851}.

Although the integration of radio-based sensing with radio information transmission may enhance the service diversity, the throughput capacity becomes a bottleneck of such ULAP systems. To tackle this issue, movable-antenna (MA) array is proposed as a key technology for the next-generation wireless communication networks, where the spatial degree of freedom of the antenna array is fully utilized to enhance the communication quality \cite{10286328}. Specifically, MA array
is capable of improving the array gain over a desired direction, while reducing the array gain toward undesired
directions that may lead to interference \cite{10278220}.

Previous works have verified the advantages of the UAV-enabled communication systems \cite{8663615, 8489918, 9604506, 9273074}.
The energy consumption of a UAV-enabled communication system is minimized in \cite{8663615}, where the throughput of the served ground users can be ensured by jointly optimizing the maneuver of the UAV, the time allocation among ground users, and the total mission completion time.
In \cite{8489918}, a UAV-enabled wireless powered communication network is studied, where the system throughput is optimized by the joint design of the transmit power, trajectory, and time allocation.
In \cite{9604506}, the minimization of the age of information (AoI) for all users involves joint optimization of the data collection time allocation, the transmit power, and the UAV trajectory.
Moreover, the uplink transmission is investigated in a UAV-enabled communication system \cite{9273074}, where the throughput of the UAV is maximized under probabilistic LoS channel.

By embracing ISAC into wireless networks, several prior works have investigated the system design, optimization, and deployment \cite{10380513, 10462908, 10233771, 10382465, 9916163}.
The coordinated transmit beamforming for information and sensing is jointly exploited in a multi-antenna multi-base station ISAC system \cite{10380513}.
A downlink and uplink cooperative ISAC system is investigated in \cite{10462908}, where a series of parameters is accurately estimated.
The average AoI of all users is minimized in \cite{10233771}, by jointly optimizing the UAV trajectory, the total time, and the sensing policy.
In \cite{10382465}, the simultaneous wireless information and power transfer technology is introduced into an ISAC system, where the trade-off among the estimated Cramér-Rao bound, achievable data rate, and energy harvesting is well investigated.
A UAV-ISAC system is studied in \cite{9916163}, where the system throughput is optimized, benefited from the mobility of the UAV's deployment.

To improve the beam gain, the combination of MA array and wireless networks is investigated recently \cite{10382559, 10504625, liu2024}.
In \cite{10382559}, by optimizing the antenna position of the MA array, the multi-beamforming gain with a linear MA array is maximized, while the interference is suppressed.
A weighted sum rate maximization problem is considered in an MA array empowered wireless system, where the computation time is reduced by approximately 30\% \cite{10504625}.
Besides, the maneuver, beamforming and positions of the MA array aided UAV-enabled system is jointly optimized in \cite{liu2024}, where the total achievable data rate is maximized to enhance the quality of services for users.

The existing ISAC system over ULAP has not been fully exploited, while the throughput capacity is still a severe bottleneck that constraints the overall system performance.
Different from the previous works and motivated by the advantages in flexible beamforming of the MA array technology, we investigate the throughput maximization problem in an MA array empowered ISAC system over ULAP for supporting LAE applications.
We jointly optimize the transmit information and sensing beamforming, and the antenna positions of the MA array using an efficient alternating optimization (AO)-based algorithm. Numerical results demonstrate the significant improvement of our proposed scheme compared with the fixed position array scheme and the random position array scheme, in terms of the achievable data rate and beamforming gain.
\textit{Organizations:} The rest of this paper is organized as follows. The system model and problem formulation are presented in Section II. Then, the AO-based algorithm is introduced in Section III. Next, Section IV shows the numerical results. Finally, Section V concludes this paper.

\textit{Notations:} The notation mentioned in this paper is introduced below.
$ {{\mathbb{C}}^{M\times N}} $ denotes the $ M \times N $ complex matrix $ \mathbb{C} $. $ {\cal C}{\cal N}(\mu ,{\sigma ^2}) $ denotes the circularly symmetric complex Gaussian distribution with $ \mu $ mean and $ {\sigma ^2} $ variance.
$ \mathrm{j} $ represents the imaginary unit, where $ {\mathrm{j}^2} = -1 $. For a generic matrix $ {\boldsymbol{G}} $, $ {{\boldsymbol{G}}^{\mathsf{H}}} $, $ {{\boldsymbol{G}}^{\mathsf{T}}} $, $ \mathsf{Tr}({\boldsymbol{G}}) $, and $\mathsf{rank}({\boldsymbol{G}})$ denote the conjugate transpose, transpose, trace, and rank of $ {\boldsymbol{G}} $, respectively.
For a vector $ {\boldsymbol{v}} $, $ \Vert {\boldsymbol{v}}\Vert $ denotes the Euclidean norm.

\begin{figure}[!htbp]
	\centering
	\includegraphics[width=0.728\linewidth]{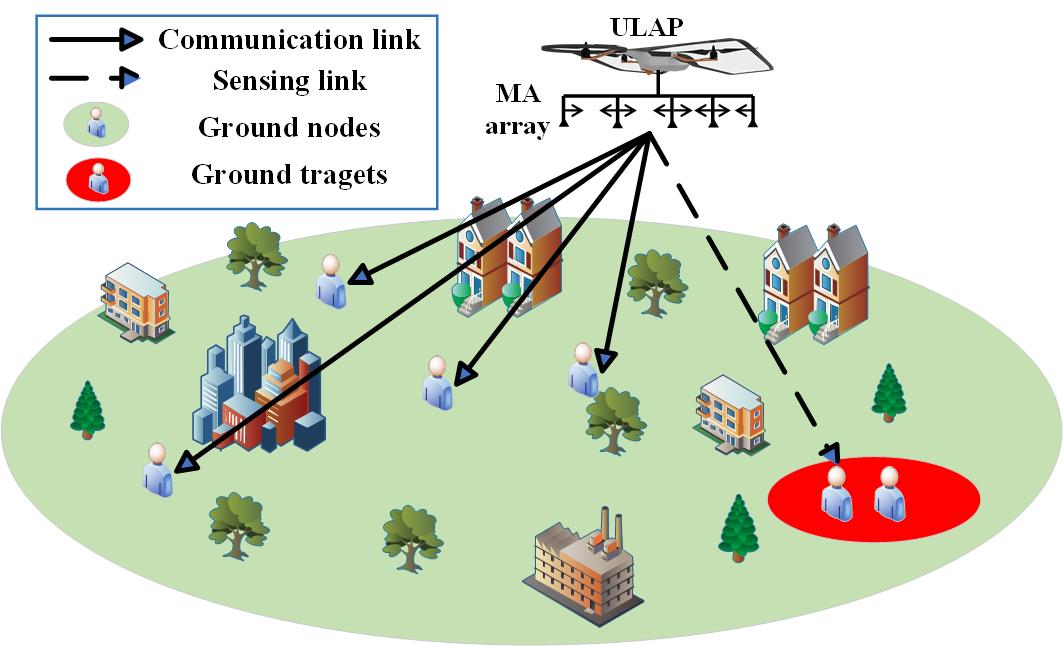}
	\caption{The MA array empowered ISAC over ULAP system.}
	\label{fsm}
\end{figure}
\begin{figure*}[!t]
\normalsize
\setcounter{mytempeqncnt}{\value{equation}}
\setcounter{equation}{5}
\begin{equation}{ \small
\label{sinr}
\begin{aligned}
\gamma_k (\boldsymbol{a}_k(\boldsymbol{x}(t)), \boldsymbol{w}_k(t),\boldsymbol{S}(t)) = \frac{ \left \vert h_k(t)\boldsymbol{a}_k^{\mathsf{H}}(t)
     \boldsymbol{w}_k (t) \right \vert^{2} }
    {\sum_{l\in\mathcal{K}^-\backslash\{k\} } \vert 
    h_k(t)\boldsymbol{a}_k^{\mathsf{H}}(t) \boldsymbol{w}_l(t) \vert^{2} + h_k^2 (t) \boldsymbol{a}_k^{\mathsf{H}} (t) \boldsymbol{S} (t) \boldsymbol{a}_k (t) + \sigma_k^2(t)},
\end{aligned} }
\end{equation}
\setcounter{equation}{\value{mytempeqncnt}}
\end{figure*}
\begin{figure*}[!t]
\normalsize
\setcounter{mytempeqncnt}{\value{equation}}
\setcounter{equation}{6}
\begin{equation} { \small
\label{rate}
\begin{aligned}
    r_k(\boldsymbol{a}_k(\boldsymbol{x}(t)), \boldsymbol{w}_k(t),\boldsymbol{S}(t)) = \log_2 (1+\gamma_k (\boldsymbol{a}_k(\boldsymbol{x}(t)), \boldsymbol{w}_k(t),\boldsymbol{S}(t))).
\end{aligned}}
\end{equation}
\setcounter{equation}{\value{mytempeqncnt}}
\hrulefill
\end{figure*}

\section{System model and Problem Formulation}

\subsection{System Model}
As illustrated in Fig.~\ref{fsm}, we consider an ISAC system deployed in the ULAP with the MA array to provide radar sensing towards ground target, and downlink communication service for $K$ single-antenna ground nodes (GNs). The set of GNs is denoted as $\mathcal{K}^- \triangleq \{1, \cdots, K\}$. We denote the ground target as the $(K+1)$-th GN, and the overall set of GNs as $\mathcal{K} \triangleq \{1, \cdots, K+1\}$. We consider a time interval with $\mathcal{T}$ seconds, which is divided into $T$ slots. Therefore, each time slot has a time duration $\tau = \frac{\mathcal{T}}{T}$. We assume that all GNs and the ULAP are fixed,
and the distance between the ULAP and the $k$-th GN is denoted by $d_k$.

The ULAP is equipped with a linear MA array with $M$ antennas \cite{10286328}, where the set of the MA antennas is given by $\mathcal{M} \triangleq \{1, \cdots, M\}$. Besides, the positions of the MA array at time slot $t$ is expressed by $\boldsymbol{x}(t) = [x_1(t),\cdots,x_M(t)]^{\mathsf{T}}$, where $x_m (t)$ is the $m$-th MA at time slot $t$. Thus, the steering vector of the MA array can be expressed as
\begin{equation}{ \small
\begin{split}
    \boldsymbol{a}_k(\boldsymbol{x}(t)) = \left [
    e^{\mathrm{j}\frac{2\pi}{\lambda}x_1(t)\cos\theta_k},
    \cdots,
    e^{\mathrm{j}\frac{2\pi}{\lambda}x_M(t)\cos\theta_k}
    \right ]^{\mathsf{T}}, \label{steering vector}
\end{split} }
\end{equation}
where $\theta_k$ represents the steering angle of the $k$-th GN and $\lambda$ denotes the wavelength.
The positions of all antennas can be adjusted in the range of $[0,L]$, where $L$ is the maximum adjustable distance.
To avoid coupling effect, any two antennas should keep a minimum distance, $d_{\min}$, i.e., $\vert x_{i} (t) - x_{j} (t) \vert \geq d_{\min}$, where $\{i, j \} \in \mathcal{M}$, $\ i \neq j$, and $\forall t$.

We consider the air-to-ground (A2G) channel modeling consisting of large-scale fading and small-scale fading, respectively. Let $g_{k} = \frac{h_0}{d^2_k}$ denote the large-scale fading at the $k$-th GN, and $\tilde{h}_{k}(t) = \sqrt{ \frac{\kappa}{\kappa+1} } + \sqrt{ \frac{1} {\kappa+1} }\tilde{g}_{k}(t)$ denote the small-fading at the $k$-th GN, where $h_0$ is a constant channel gain for a reference distance $1 \rm{m}$, $\kappa$ is the Rician factor, and $\tilde{g}_k(t) \sim \mathcal{C}\mathcal{N}(0,1)$ represents the channel noise. Thus, the channel gain of the A2G channel between the ULAP and the $k$-th GN is given by
\begin{equation} {\small
\begin{split}
    h_k(t) = \sqrt{g_{k}} \tilde{h}_{k}(t)=\sqrt{\frac{h_0}{d^2_k}}\left (\sqrt{ \frac{\kappa}{\kappa+1} } + \sqrt{ \frac{1} {\kappa+1} }\tilde{g}_{k}(t)
    \right ).
\end{split}}
\end{equation}

The ULAP transmits the information signals to $K$ GNs, and also transmits the dedicated sensing signals to the ground target. The information signal transmitted to the $k$-th GN is denoted as $s_k(t), k \in \mathcal{K}^-$.
The sensing signal transmitted to the target is denoted as $\boldsymbol{s}(t) \in \mathbb{C}^{M\times 1}$, which is an independently generated random vector with zero mean and covariance matrix $\boldsymbol{S}(t)=\mathbb{E}(\boldsymbol{s}(t)\boldsymbol{s}(t)^{\mathsf{H}})\succeq\boldsymbol{0}$.
Note that the multi-beam transmission for the radar sensing signal is considered in this paper. According to \cite{9916163}, $\boldsymbol{S}(t)$ is of general rank as $\mathsf{rank}(\boldsymbol{S}(t)) = M_s$, where $0 \leq M_s \leq M$. The forming of radar signal beams set is derived by the eigenvalue decomposition of $\boldsymbol{S}(t)$.
Let $\boldsymbol{w}_k(t)$ represent the transmit beamforming vector for communicating with the $k$-th GN.
Thus, the transmitted signal by the ULAP at time slot $t$ can be expressed as
\begin{equation} {\small
    \boldsymbol{z}(t) = \sum_{k=1}^K\boldsymbol{w}_k(t)s_k(t) + \boldsymbol{s}(t). }
\end{equation}
The average transmit power of the ULAP is given by $\mathbb{E}(\Vert \boldsymbol{z}(t) \Vert^2) = \sum_{k= 1}^{K} \Vert \boldsymbol{w}_k(t)) \Vert^2 + \mathsf{tr}(\boldsymbol{S}(t))$, which cannot exceed the maximum transmit power of the ULAP $P_{\max}$, i.e.,
\begin{equation} {\small
    \sum_{k= 1}^{K} \Vert \boldsymbol{w}_k(t)) \Vert^2 + \mathsf{tr}(\boldsymbol{S}(t)) \leq P_{\max},\quad \forall t.}
\end{equation}
Accordingly, the received signal by the $k$-th GN at time slot $t$ can be expressed as
\begin{equation} {\small
    y_k(t)=h_k(t)\boldsymbol{a}_k^{\mathsf{H}}(\boldsymbol{x}(t)) \boldsymbol{z}(t) + n_k(t),\quad \forall k \in \mathcal{K}^-,}
\end{equation}
where $n_k(t)$ represents the additive white Gaussian
noise (AWGN) of the $k$-th GN.
Meanwhile, the received signal-to-interference-plus-noise ratio (SINR) of the $k$-th GN at time slot $t$ can be derived as Eq. (\ref{sinr}),
where $\sigma_k^2(t)$ denotes the noise power. The achievable data rate transmitted by the ULAP towards the $k$-th user at time slot $t$ can be given by Eq. (\ref{rate}).

On the other hand, to improve the sensing and communication performance,
we utilize both the communication signal and the sensing signal for dedicated sensing \cite{9916163, 9124713, 10086626}. Therefore, the transmit beampattern gain toward the target is
\begin{equation}  {\small
\begin{split}
  \mathcal{B}(t)&=
   \mathbb{E}\left(
        \left \vert \boldsymbol{a}_{K+1}^{\mathsf{H}}(\boldsymbol{x}(t))   \boldsymbol{z}(t)
        \right \vert^2
    \right)\\
    &= 
    \boldsymbol{a}_{K+1}^{\mathsf{H}}(\boldsymbol{x}(t))\boldsymbol{\mathcal{W}}(t)\boldsymbol{a}_{K+1}(\boldsymbol{x}(t)),
\end{split} }
\addtocounter{equation}{2}
\end{equation}
where $\boldsymbol{\mathcal{W}}(t) = 
        \sum_{k=1}^K\boldsymbol{w}_k(t)\boldsymbol{w}_k^{\mathsf{H}}(t) + \boldsymbol{S}(t)$.

\begin{figure*}[!t]
\normalsize
\setcounter{mytempeqncnt}{\value{equation}}
\setcounter{equation}{16}
\begin{equation} {\small
\label{rate2}
\begin{split}
\Tilde{r}^{\rm{}}_k (\boldsymbol{W}_k(t),\boldsymbol{S}(t)) =& \log_2\left(1+
        \frac{ \mathsf{tr}\left( h_k^2(t)\boldsymbol{a}_k(t)\boldsymbol{a}_k^{\mathsf{H}}(t)
     \boldsymbol{W}_k (t)  \right) }
    {\sum_{l\in\mathcal{K}^-\backslash\{k\} } \mathsf{tr}\left( h_k^2(t)\boldsymbol{a}_k(t)\boldsymbol{a}_k^{\mathsf{H}}(t)
     \boldsymbol{W}_l (t)  \right) + 
     \mathsf{tr}\left(h_k^2 (t) \boldsymbol{a}_k (t)\boldsymbol{a}_k^{\mathsf{H}} (t) \boldsymbol{S} (t) \right) + \sigma_k^2(t)}
\right)\\
=& \log_2\left(\sum_{l\in\mathcal{K}^- } \mathsf{tr}\left( h_k^2(t)\boldsymbol{a}_k(t)\boldsymbol{a}_k^{\mathsf{H}}(t)
     \boldsymbol{W}_l (t)  \right) + 
     \mathsf{tr}\left(h_k^2 (t) \boldsymbol{a}_k (t)\boldsymbol{a}_k^{\mathsf{H}} (t) \boldsymbol{S} (t) \right) + \sigma_k^2(t)
\right)\\
&-  \log_2\left(\sum_{l\in\mathcal{K}^-\backslash\{k\} } \mathsf{tr}\left( h_k^2(t)\boldsymbol{a}_k(t)\boldsymbol{a}_k^{\mathsf{H}}(t)
     \boldsymbol{W}_l (t)  \right) + 
     \mathsf{tr}\left(h_k^2 (t) \boldsymbol{a}_k (t)\boldsymbol{a}_k^{\mathsf{H}} (t) \boldsymbol{S} (t) \right) + \sigma_k^2(t)\right).
\end{split} }
\end{equation}
\setcounter{equation}{\value{mytempeqncnt}}
\end{figure*}
\begin{figure*}[!t]
\normalsize
\setcounter{mytempeqncnt}{\value{equation}}
\setcounter{equation}{18}
\begin{equation}{\small{
\label{frakt}
\begin{aligned}
    \boldsymbol{\mathcal{L}}_k(t) = 
    \frac{h_k^2(t) \boldsymbol{a}_k(t) \boldsymbol{a}_k^{\mathsf{H}}(t)}
    {\ln{2}\left(
        \sum_{l\in\mathcal{K}^-\backslash\{k\} }
        \mathsf{tr} \left(h_k^2(t) \boldsymbol{a}_k(t) \boldsymbol{a}_k^{\mathsf{H}}(t) \boldsymbol{W}_l^{(i)}(t)\right)
        + \mathsf{tr} \left(h_k^2(t) \boldsymbol{a}_k(t) \boldsymbol{a}_k^{\mathsf{H}}(t) \boldsymbol{S}^{(i)}(t)\right) +\sigma_k^2(t)
    \right)}.
\end{aligned}}}
\end{equation}
\setcounter{equation}{\value{mytempeqncnt}}
\hrulefill
\end{figure*}
\subsection{Problem Formulation}
In this paper, we aim to maximize the communication rate while ensuring the dedicated sensing performance.
By jointly optimizing the transmit information beamforming, $\boldsymbol{W} \triangleq \{ \boldsymbol{w}_k(t), \forall k, \forall t\}$, sensing beamforming, $\boldsymbol{S} \triangleq \{ \boldsymbol{S}(t), \forall t\}$, and the positions of MA array $\boldsymbol{X} \triangleq \{ \boldsymbol{x}(t), \forall t\}$,
the overall data rate maximization problem can be formulated as
\begin{subequations} { \small
\begin{flalign}
 (\textbf{P1}):\ \max_{\boldsymbol{W},\boldsymbol{S}, \boldsymbol{X}} \ \ & \sum_{t= 1}^{T}\sum_{k= 1}^{K}  r_k(\boldsymbol{a}_k(\boldsymbol{x}(t)), \boldsymbol{w}_k(t),\boldsymbol{S}(t))  \nonumber\\
 {\rm{s.t.}}  \ \ 
 & \{x_m[n]\}_{m=1}^M \in [0,L],\ \forall t,\label{p1a}\\
 & \vert x_{i} (t) - x_{j} (t) \vert \geq d_{\min},\nonumber\\
 & \quad\quad\quad \{i, j \} \in \mathcal{M},\ i \neq j, \ \forall t,\label{p1b}\\
 &\sum_{k= 1}^{K} \Vert \boldsymbol{w}_k(t)) \Vert^2 + \mathsf{tr}(\boldsymbol{S}(t)) \le P_{\max},\ \forall t,   \label{p1c}\\
 &\mathcal{B}(t) \geq \Lambda_{\mathrm{th}},\ \forall t,   \label{p1d}
\end{flalign}}\end{subequations} where $\Lambda_{\mathrm{th}}$ is the threshold of the beampattern gain that ensures the performance of dedicated sensing signal.
(\ref{p1a}) and (\ref{p1b}) represent the position constraints of the MA array, (\ref{p1c}) and (\ref{p1d}) denote the ULAP power limitation and the sensing beamforming gain requirement, respectively.
Problem (\textbf{P1}) is a non-convex optimization problem due to the coupling of the information and sensing beamforming $\boldsymbol{W}$, $\boldsymbol{S}$, and the positions of the MA array $\boldsymbol{X}$.
To tackle this issue, we will introduce an AO-based algorithm in the next section.

\section{Joint Optimization of the Transmit Information and Sensing Beamforming, and Positions of the MA Array}

\subsection{MA Array Position Optimization}
Given the beamforming $\boldsymbol{W}$ and $\boldsymbol{S}$, the problem (\textbf{P1}) can be simplified as
\begin{subequations} {\small
\begin{flalign}
 (\textbf{P2}):\ \max_{\boldsymbol{X}} \quad & \sum_{t= 1}^{T}\sum_{k= 1}^{K} {r}^{\rm{}}_k (\boldsymbol{a}_k(\boldsymbol{x}(t))) \nonumber\\
 {\rm{s.t.}}  \quad & \text{(\ref{p1a}), (\ref{p1b}) and (\ref{p1d})}. \nonumber
\end{flalign} }
\end{subequations}

We utilize the particle swarm optimization (PSO)-based method to solve (\textbf{P2}). In particular, $S$ searching solutions of (\textbf{P2}) are randomly selected as a particle swarm from the feasible set, where each is a particle of PSO mechanism. We denote solution of the $s$-th particle during $i$-th iteration as $\boldsymbol{x}^{s}_{(i)} = [x^{s}_{1,(i)},\cdots,x^{s}_{M,(i)}]^{\mathsf{T}}$. Then, each particle updates its local optimal solution iteratively based on its own exploitation and the sharing solutions from the swarm.
Finally, all particles will converge to the global optimal solution $\boldsymbol{x}^{*} = [x_1^{*},\cdots,x_M^{*}]^{\mathsf{T}}$.

The local optimal solution $\boldsymbol{x}^{s*}_{(i)}$ of the $s$-th particle during the $i$-th iteration is given by
\begin{equation} { \small
   \boldsymbol{x}^{s*}_{(i)} = {\arg\max}_{\boldsymbol{x}^{s}\in\{\boldsymbol{x}^{s*}_{(1)},\cdots,\boldsymbol{x}^{s*}_{(i)} \}} \sum_{t= 1}^{T}\sum_{k= 1}^{K} {r}^{\rm{}}_k (\boldsymbol{a}_k(\boldsymbol{x}(t))).} \label{pbest}
\end{equation}
Besides, the global optimal solution $\boldsymbol{x}^{*}_{(i)}$ from the swarm is given by
\begin{equation} { \small
   \boldsymbol{x}^{*}_{(i)} = {\arg\max}_{\boldsymbol{x}^{s}\in\{\boldsymbol{x}^{1*}_{(i)},\cdots,\boldsymbol{x}^{S*}_{(i)} \}} \sum_{t= 1}^{T}\sum_{k= 1}^{K} {r}^{\rm{}}_k (\boldsymbol{a}_k(\boldsymbol{x}(t))).} \label{gbest}
\end{equation}

The exploitation of the $s$-th particle is updated by
\begin{equation}  {\small
\begin{split}
       \mathcal{V}^{s}_{(i+1)} = \omega \mathcal{V}^{s}_{(i)} &+ \mathcal{C}\mathcal{R}^{s}_{1,(i)} (\boldsymbol{x}^{s*}_{(i)} - \boldsymbol{x}^{s}_{(i)})  \\&+\mathcal{S}\mathcal{R}^{s}_{2,(i)} (\boldsymbol{x}^{*}_{(i)}-\boldsymbol{x}^{s}_{(i)}), \label{velocity}
\end{split} }
\end{equation}
where $\mathcal{V}^{s}_{(i)}$ denotes the update velocity of the $s$-th particle during $i$-th iteration, $\omega$ represents the inertia weight, $\mathcal{C}$ and $\mathcal{S}$ denote the cognitive parameter and social parameter, respectively, which are both positive constants. $\mathcal{R}^{s}_{1,(i)} \sim \mathcal{U}(0,1)$ and $\mathcal{R}^{s}_{2,(i)} \sim \mathcal{U}(0,1)$ are two random variables following uniform distribution. Hence, the solution is updated by
\begin{equation}  {\small
   \boldsymbol{x}^{s}_{(i+1)} = \boldsymbol{x}^{s}_{(i)} + \varpi \mathcal{V}^{s}_{(i+1)}, }\label{position}
\end{equation}
where $\varpi$ is a constant for update control. To ensure that the solutions obtained through PSO satisfy the constraints of (\textbf{P2}), we make some modifications to the PSO algorithm, as illustrated in Algorithm \ref{Alg}, where $i_{\max}$ is the maximum iteration number of PSO. By applying these methods, we can obtain the solutions that adhere to the constraints. In addition, the objective value of (\textbf{P2}) is bounded, ensuring the convergence of Algorithm \ref{Alg}.

\begin{algorithm} \scriptsize
\caption{Modified PSO for Solving (\textbf{P2})}
\label{Alg}
\begin{algorithmic}
  \REQUIRE { The initial swarm with $S$ particles, the max iteration number $i_{\max}$; }\\
  \FOR { $s$ = 1 to $S$ }  
  \STATE Randomly initialize the position $\boldsymbol{x}^{s}_{(0)}$ within the given bounds;\\
  \REPEAT
  \STATE Check if particle $s$ satisfies constraint (\ref{p1b}), else reconstruct $s$ through $x_{m+1,(0)}^{s} = x_{m,(0)}^{s} + d_{\min}$;\\ 
  \STATE Check if particle $s$ satisfies constraint (\ref{p1d}), else randomly initialize the position $\boldsymbol{x}^{s}_{(0)}$;\\ 
  \UNTIL {the particle $s$ satisfies (\ref{p1b}) and (\ref{p1d})};\\
  \STATE Randomly initialize the velocity $\mathcal{V}^{s}_{(0)}$;\\
  \ENDFOR \\
  
  \STATE Set the particle's own optimal position $\boldsymbol{x}^{s*}_{(0)} = \boldsymbol{x}^{s}_{(0)}$ and the global optimal position $\boldsymbol{x}^{*}_{(0)} = \boldsymbol{x}^{s}_{(0)}$;\\
  \FOR{ $i = 1$ to $i_{\max}$}         
  \STATE  Update the inertia weight $\omega$ according to $\omega = 0.99 \omega $;\\
  \FOR{$s = 1$ to $S$}
    \STATE Update the particle $s$ according to (\ref{velocity}) and (\ref{position});\\
  \REPEAT
  \STATE Check if particle $s$ satisfies constraint (\ref{p1b}), else reconstruct $s$ through $x_{m+1}^{s,(i)} = x_{m}^{s,(i)} + d_{\min}$;\\ 
  \STATE Check if particle $s$ satisfies constraint (\ref{p1d}), else randomly initialize the position $\boldsymbol{x}^{s,(i)}$;\\ 
  \UNTIL {the particle $s$ satisfies (\ref{p1b}) and (\ref{p1d})};\\
  \STATE Evaluate the fitness of particle $s$;\\
  \STATE Update the particle's own optimal position according to (\ref{pbest});\\
  \STATE Update the global optimal position according to (\ref{gbest});\\
  \ENDFOR
  \ENDFOR	
  
  \ENSURE {the global optimal solution $\boldsymbol{x}^{*}_{(i_{\max})}$}\\
 \end{algorithmic}
\end{algorithm}

\subsection{Information and Sensing Beamforming Optimization}
Given the position of MA array $\boldsymbol{X}$, the problem (\textbf{P1}) can be reformulated as
\begin{subequations} { \small
\begin{flalign}
 (\textbf{P3}):\ \max_{\boldsymbol{W},\boldsymbol{S}} \quad & \sum_{t= 1}^{T}\sum_{k= 1}^{K} {r}^{\rm{}}_k (\boldsymbol{w}_k(t),\boldsymbol{S}(t)) \nonumber\\
 {\rm{s.t.}}  \quad & \text{(\ref{p1c}) and (\ref{p1d})}. \nonumber
\end{flalign}}\end{subequations}Let $\boldsymbol{W}_k(t) \triangleq \boldsymbol{w}_k(t)\boldsymbol{w}_k^{\mathsf{H}}(t)$, where $\boldsymbol{W}_k(t) \succeq \boldsymbol{0}$, and $\mathsf{rank}(\boldsymbol{W}_k(t))\leq 1$. Substituting $\boldsymbol{w}_k(t)\boldsymbol{w}_k^{\mathsf{H}}(t)$ by $\boldsymbol{W}_k(t)$, problem (\textbf{P3}) is reformulated as
\begin{subequations} {\small
\begin{flalign}
 (\textbf{P4}):\ \max_{\boldsymbol{W}_k,\boldsymbol{S}} \quad & \sum_{t= 1}^{T}\sum_{k= 1}^{K} \Tilde{r}^{\rm{}}_k (\boldsymbol{W}_k(t),\boldsymbol{S}(t)) \nonumber\\
 {\rm{s.t.}}  \quad & \text{(\ref{p1d})}, \nonumber\\
 &\sum_{k= 1}^{K} \mathsf{tr}(\boldsymbol{W}_k(t)) + \mathsf{tr}(\boldsymbol{S}(t)) \le P_{\max},\ \forall t,   \label{p4a}\\
 &\mathsf{rank}(\boldsymbol{W}_k(t))\leq 1,\ \forall t,\label{p4b}
\end{flalign}}\end{subequations}where $\Tilde{r}^{\rm{}}_k (\boldsymbol{W}_k(t),\boldsymbol{S}(t))$ is expressed as Eq.(\ref{rate2}).
The objective of (\textbf{P4}) is still non-concave, thus we approximate it by
utilizing the first-order Taylor expansion based successive convex approximation (SCA) on the second term of Eq.(\ref{rate2}).
Hence, the approximation of Eq.(\ref{rate2}) can be given by
\begin{equation}\label{p4lb}{\small
\begin{split}
    &\Tilde{r}^{\rm{}}_k (\boldsymbol{W}_k(t),\boldsymbol{S}(t))\\
    \geq& \log_2\left(\sum_{l\in\mathcal{K}^- } \mathsf{tr}\left( h_k^2(t)\boldsymbol{a}_k(t)\boldsymbol{a}_k^{\mathsf{H}}(t)
     \boldsymbol{W}_l (t)  \right) + \right.\\
     &\quad\quad\quad \left.\mathsf{tr}\left(h_k^2 (t) \boldsymbol{a}_k (t)\boldsymbol{a}_k^{\mathsf{H}} (t) \boldsymbol{S} (t) \right) + \sigma_k^2(t)
    \right. \Bigg)\\
    &- \log_2\Bigg(\sum_{l\in\mathcal{K}^-\backslash\{k\} } \mathsf{tr}\left( h_k^2(t)\boldsymbol{a}_k(t)\boldsymbol{a}_k^{\mathsf{H}}(t)
     \boldsymbol{W}_l^{(i)} (t)  \right) + \\
      &\quad\quad\quad\quad \mathsf{tr}\left(h_k^2 (t) \boldsymbol{a}_k (t)\boldsymbol{a}_k^{\mathsf{H}} (t) \boldsymbol{S}^{(i)} (t) \right) + \sigma_k^2(t) \Bigg)\\
      &- \sum_{l\in\mathcal{K}^-\backslash\{k\} } \mathsf{tr}
    \left(\boldsymbol{\mathcal{L}}_k(t)\left(\boldsymbol{W}_l (t)-\boldsymbol{W}_l^{(i)} (t)\right)\right)\\
      &- \mathsf{tr}
      \left(\boldsymbol{\mathcal{L}}_k(t)\left(\boldsymbol{S} (t)-\boldsymbol{S}^{(i)} (t)\right)\right)
      \triangleq \check{r}^{(i)}_{k,\mathrm{lb}} (\boldsymbol{W}_k(t),\boldsymbol{S}(t)),
\end{split}}
\addtocounter{equation}{1}\end{equation}
where $\boldsymbol{\mathcal{L}}_k(t)$ is defined by Eq.(\ref{frakt}).
Besides, since the right hand side of (\ref{p4lb}) is the lower bound of the objective of (\textbf{P4}), the relaxed (\textbf{P4}.$i$) under the $i$-th iteration is derived by
\begin{subequations} {\small
\begin{flalign}
 (\textbf{P4}.i):\ \max_{\boldsymbol{W}_k,\boldsymbol{S}} \quad & \sum_{t= 1}^{T}\sum_{k= 1}^{K} \check{r}^{(i)}_{k,\mathrm{lb}} (\boldsymbol{W}_k(t),\boldsymbol{S}(t)) \nonumber\\
 {\rm{s.t.}}  \quad & \text{(\ref{p1d}), (\ref{p4a}), and (\ref{p4b})}. \nonumber
\end{flalign}}
\addtocounter{equation}{1}
\end{subequations}
Note that constraint (\ref{p4b}) is still non-convex. To address this issue, we relax (\textbf{P4}.$i$) as (\textbf{P4}$^*$.$i$) through the semi-definite relaxation (SDR), which is given by
\begin{subequations} { \small
\begin{flalign}
 (\textbf{P4}^*.i):\ \max_{\boldsymbol{W}_k,\boldsymbol{S}} \quad & \sum_{t= 1}^{T}\sum_{k= 1}^{K} \check{r}^{(i)}_{k,\mathrm{lb}} (\boldsymbol{W}_k(t),\boldsymbol{S}(t)) \nonumber\\
 {\rm{s.t.}}  \quad & \text{(\ref{p1d}) and (\ref{p4a}),} \nonumber\\
 &\boldsymbol{W}_k(t) \succeq 0,\ \forall k,\ \forall t.
\end{flalign}}
\end{subequations}The problem (\textbf{P4}$^*$.$i$) is convex, thus can be solved by CVX.
\begin{proposition}
The global optimal solution to (\textbf{P4}$^*$.$i$), which is denoted as $\{\boldsymbol{\Tilde{W}}_k^{}, \forall k \in \mathcal{K}^-, \boldsymbol{\Tilde{S}} \}$, always exists and satisfies $\mathsf{rank}(\boldsymbol{\Tilde{W}}_k)=1,\forall k \in \mathcal{K}^-$.
\begin{proof}
We denote the optimal solution for (\textbf{P4}$^*$.$i$) as 
$\{\boldsymbol{W}_k^{*}, \forall k \in \mathcal{K}^-, \boldsymbol{S}^{*} \}$.
If the solution is rank-one, the SDR is tight, and the solution is also the solution of (\textbf{P4}.$i$). Otherwise, we can construct $\{\boldsymbol{\Tilde{W}}_k^{}, \forall k \in \mathcal{K}^-, \boldsymbol{\Tilde{S}}^{} \}$ as
\begin{equation} {\small
    \boldsymbol{\Tilde{W}}_k = (\Tilde{\boldsymbol{w}}_k^{} )^{\mathsf{H}}\Tilde{\boldsymbol{w}}_k^{},}\label{prop1}
\end{equation}
where $\Tilde{\boldsymbol{w}}_k^{} = \left( 
    \boldsymbol{a}_k^{\mathsf{H}} \boldsymbol{W}_k^{*}\boldsymbol{a}_k
    \right)^{-\frac{1}{2}} \boldsymbol{W}_k^{*}\boldsymbol{a}_k$,
\begin{equation} {\small
    \boldsymbol{\Tilde{S}} = \boldsymbol{S}^* + \sum_{k \in \mathcal{K}^-} \boldsymbol{W}_k^{*} - \sum_{k \in \mathcal{K}^-} \boldsymbol{\Tilde{W}}_k,}\label{prop2}
\end{equation}
where the time slot notation $(t)$ and the iteration superscript $(i)$ are omitted, for simplicity. Meanwhile, it is observed that $\boldsymbol{\Tilde{W}}_k^{} \succeq \boldsymbol{0}, \forall k \in \mathcal{K}^-$ is positive semi-definite with $\mathsf{rank}(\boldsymbol{\Tilde{W}}_k^{}) =1$. For any given vector $\boldsymbol{v} \in \mathbb{C}^M$, we have
\begin{equation} {\small
\begin{split}
    \boldsymbol{v}^{\mathsf{H}} &\left( \boldsymbol{W^*} -\boldsymbol{\Tilde{W}} \right ) \boldsymbol{v}
    = \boldsymbol{v}^{\mathsf{H}}\boldsymbol{W^*}\boldsymbol{v}-
    \left( 
    \boldsymbol{a}_k^{\mathsf{H}} \boldsymbol{W}_k^{*}\boldsymbol{a}_k
    \right)^{-1}
    \vert \boldsymbol{v}^{\mathsf{H}}\boldsymbol{W}_k^{*}\boldsymbol{a}_k\vert^2 \\
    &\geq \boldsymbol{v}^{\mathsf{H}}\boldsymbol{W^*}\boldsymbol{v} -
    \left( 
    \boldsymbol{a}_k^{\mathsf{H}} \boldsymbol{W}_k^{*}\boldsymbol{a}_k
    \right)^{-1}
    \vert \boldsymbol{v}^{\mathsf{H}}\boldsymbol{w}_k^{*}\vert^2\vert\boldsymbol{a}_k^{\mathsf{H}}\boldsymbol{w}_k^{*}\vert^2 = \boldsymbol{0}.
\end{split}} \label{prop4}
\end{equation}
From (\ref{prop2}) and (\ref{prop4}), we can derive that $\boldsymbol{\Tilde{S}} \succeq \boldsymbol{0}$ is also positive semi-definite.

By (\ref{prop1}), we can simply derive that $ \boldsymbol{a}_k^{\mathsf{H}} \boldsymbol{\Tilde{W}}_k \boldsymbol{a}_k = \boldsymbol{a}_k^{\mathsf{H}} \boldsymbol{W}_k^{*} \boldsymbol{a}_k $. By substituting (\ref{prop1}) and (\ref{prop2}) into the objective of (\textbf{P4*}.$i$), i.e., $\check{r}^{(i)}_{k,\mathrm{lb}} $, we can obtain
\begin{equation} {\footnotesize
\begin{split}
        &\check{r}^{}_{k} (\boldsymbol{\Tilde{W}}_k,\boldsymbol{\Tilde{S}})=
    \log_2\left(\sum_{l\in\mathcal{K}^- } \mathsf{tr}\left( h_k^2\boldsymbol{a}_k\boldsymbol{a}_k^{\mathsf{H}}
     \boldsymbol{\Tilde{W}}_l  \right) + 
     \mathsf{tr}\left(h_k^2  \boldsymbol{a}_k \boldsymbol{a}_k^{\mathsf{H}}  \boldsymbol{\Tilde{S}} \right) + \sigma_k^2
    \right. \Bigg)\\
    &\quad\quad- \log_2(\digamma_k ) - \sum_{l\in\mathcal{K}^- } 
    \left(\frac{h_k^2}{\ln{2}\digamma_k}\boldsymbol{a}_k^{\mathsf{H}}\left(\boldsymbol{\Tilde{W}}_l -\boldsymbol{W}_l^{} \right)\boldsymbol{a}_k\right)\\
      &\quad\quad+ \left(\frac{h_k^2}{\ln{2}\digamma_k}\boldsymbol{a}_k^{\mathsf{H}}\left(\boldsymbol{\Tilde{W}}_k -\boldsymbol{W}_k \right)\boldsymbol{a}_k\right)
      -
      \left(\frac{h_k^2}{\ln{2}\digamma_k}\boldsymbol{a}_k^{\mathsf{H}}\left(\boldsymbol{\Tilde{S}} -\boldsymbol{S} \right)\boldsymbol{a}_k\right)\\
      &=\log_2\left(\sum_{l\in\mathcal{K}^- } \mathsf{tr}\left( h_k^2\boldsymbol{a}_k\boldsymbol{a}_k^{\mathsf{H}}
     \boldsymbol{{W}}_l^*  \right) + 
     \mathsf{tr}\left(h_k^2  \boldsymbol{a}_k \boldsymbol{a}_k^{\mathsf{H}}  \boldsymbol{{S}}^* \right) + \sigma_k^2
    \right. \Bigg)\\
    &\quad\quad- \log_2(\digamma_k ) - \sum_{l\in\mathcal{K}^- \backslash \{ k\} } 
    \left(\frac{h_k^2}{\ln{2}\digamma_k}\boldsymbol{a}_k^{\mathsf{H}}\left(\boldsymbol{{W}}_l^* -\boldsymbol{W}_l^{} \right)\boldsymbol{a}_k\right)\\
      &\quad\quad-
      \left(\frac{h_k^2}{\ln{2}\digamma_k}\boldsymbol{a}_k^{\mathsf{H}}\left(\boldsymbol{{S}}^* -\boldsymbol{S} \right)\boldsymbol{a}_k\right) = \check{r}^{}_{k} (\boldsymbol{{W}}_k^*,\boldsymbol{{S}}^*),
\end{split} }\label{prop3}
\end{equation}
where
\begin{equation}
{\small
\begin{split}
\digamma_k = 
    \sum_{l\in\mathcal{K}^-\backslash\{k\} } \mathsf{tr}\left( h_k^2\boldsymbol{a}_k\boldsymbol{a}_k^{\mathsf{H}}
     \boldsymbol{W}_l^{}\right) + \mathsf{tr}\left(h_k^2 \boldsymbol{a}_k (t)\boldsymbol{a}_k^{\mathsf{H}} \boldsymbol{S}^{}  \right) + \sigma_k^2(t).
\end{split} }
\end{equation}
Eq.(\ref{prop3}) demonstrates that the solution $\{\boldsymbol{\Tilde{W}}_k^{}, \forall k \in \mathcal{K}^-, \boldsymbol{\Tilde{S}}^{} \}$ is also the optimal solution to (\textbf{P4}.$i$).
\end{proof}
\end{proposition}
Accordingly, the SDR of (\textbf{P4}$^*$.$i$) is tight, then the solution of (\textbf{P4}$^*$.$i$), i.e., $\{\boldsymbol{\Tilde{W}}_k^{}, \forall k \in \mathcal{K}^-, \boldsymbol{\Tilde{S}}^{} \}$ is also the solution to (\textbf{P4}.$i$). With the non-decreasing nature of the objective in (\textbf{P4}$^*$.$i$) and (\textbf{P4}.$i$), the convergence can be satisfied. 

\subsection{Convergence and Complexity Analysis}
The overall AO-based algorithm for solving (\textbf{P1}) is summarized in Algorithm \ref{Alg2}.
Since the objective of (\textbf{P1}) is bounded and each variable is constrained, the objective value increases progressively with each iteration.
Therefore, Algorithm \ref{Alg2} is guaranteed to converge.
The complexity is analyzed as follows. Solving ($\textbf{P2}$) involves a complexity of $\mathcal{O} (i_{\max}S^{2}T)$. In addition, $\boldsymbol{W}$ can be obtained by solving ($\textbf{P4}^{*}.i$) with a complexity of $\mathcal{O}(j_{\max}(T(K+1)M^{2})^{3.5}\log(\epsilon_{\epsilon^*}^{-1})),$ where $\epsilon$ is the iteration accuracy, and $j_{\max}$ is the iteration number for solving ($\textbf{P4}^{*}.i$). Therefore, the total complexity of Algorithm \ref{Alg2} is $\mathcal{O}(i_{\max}(i_{\max}S^{2}T + j_{\max}(T(K+1)M^{2})^{3.5}\log(\epsilon_{\epsilon^*}^{-1}))).$
\begin{algorithm} \scriptsize
\caption{Overall AO-based Algorithm for Solving ($\textbf{P1}$)}
\label{Alg2}
\begin{algorithmic}
	
\REQUIRE {The initial $\boldsymbol{W}$, $\boldsymbol{S}$, and $\boldsymbol{X}$;}\\
\FOR{$i = 1$ to $i_{\max}$}
\STATE Update $\boldsymbol{X}$ by solving ($\textbf{P2}$) with the Algorithm \ref{Alg}; \\
\FOR{$j = 1$ to $j_{\max}$}
\STATE Update $\boldsymbol{W}$, $\boldsymbol{S}$ by solving ($\textbf{P4}^{*}.i$);
\STATE Set $j=j+1$; \\
\ENDFOR
\STATE Set $i=i+1$; \\
\ENDFOR

\ENSURE {$\boldsymbol{W}^{*}$, $\boldsymbol{S}^{*}$, and $\boldsymbol{X}^{*}$.}\\
\end{algorithmic}
\end{algorithm}

\section{Numerical Results}
In this section, we present numerical results to validate the effectiveness of the proposed scheme. In the simulation, we consider an area of $500 \mathrm{m} \times 500 \mathrm{m} $ with $K = 5$ users.
The minimum distance between adjacent antenna of the MA array is $d_{\min} = \lambda /2$. The range of positions of all antennas is $L = 10 \lambda$. The number of antennas at the ULAP is $M = 6$. The maximum transmit power is $P_{\max} = 1 \rm{W}$. The noise power at each receiver is $\sigma_k^2 = -110 \rm{dBm}$. The constant channel gain $ h_0 = -60 \rm{dB}$. The altitude of ULAP is $H = 50 \rm{m}$. The time interval $ \mathcal{T} = 10 \rm{s}$ and the total time slot $T = 10$.
The following benchmark schemes are considered for comparisons.
1) \textit{Fixed-position antenna array} (FPA) \cite{10382559} which optimizes the transmit information and sensing beamforming $\boldsymbol{w}_k$, and $\boldsymbol{S}$, while the positions of antenna array are fixed.
2) \textit{Random-position antenna array} (RPA) \cite{10508198}. In this scheme, the positions of antenna array are randomly generated while satisfying the constraints (\ref{p1a}) and (\ref{p1b}). Moreover, the transmit information and sensing beamforming $\boldsymbol{w}_k$, and $\boldsymbol{S}$ are also optimized.

\begin{figure}[!htbp]
	\centering
	\includegraphics[width=0.578\linewidth]{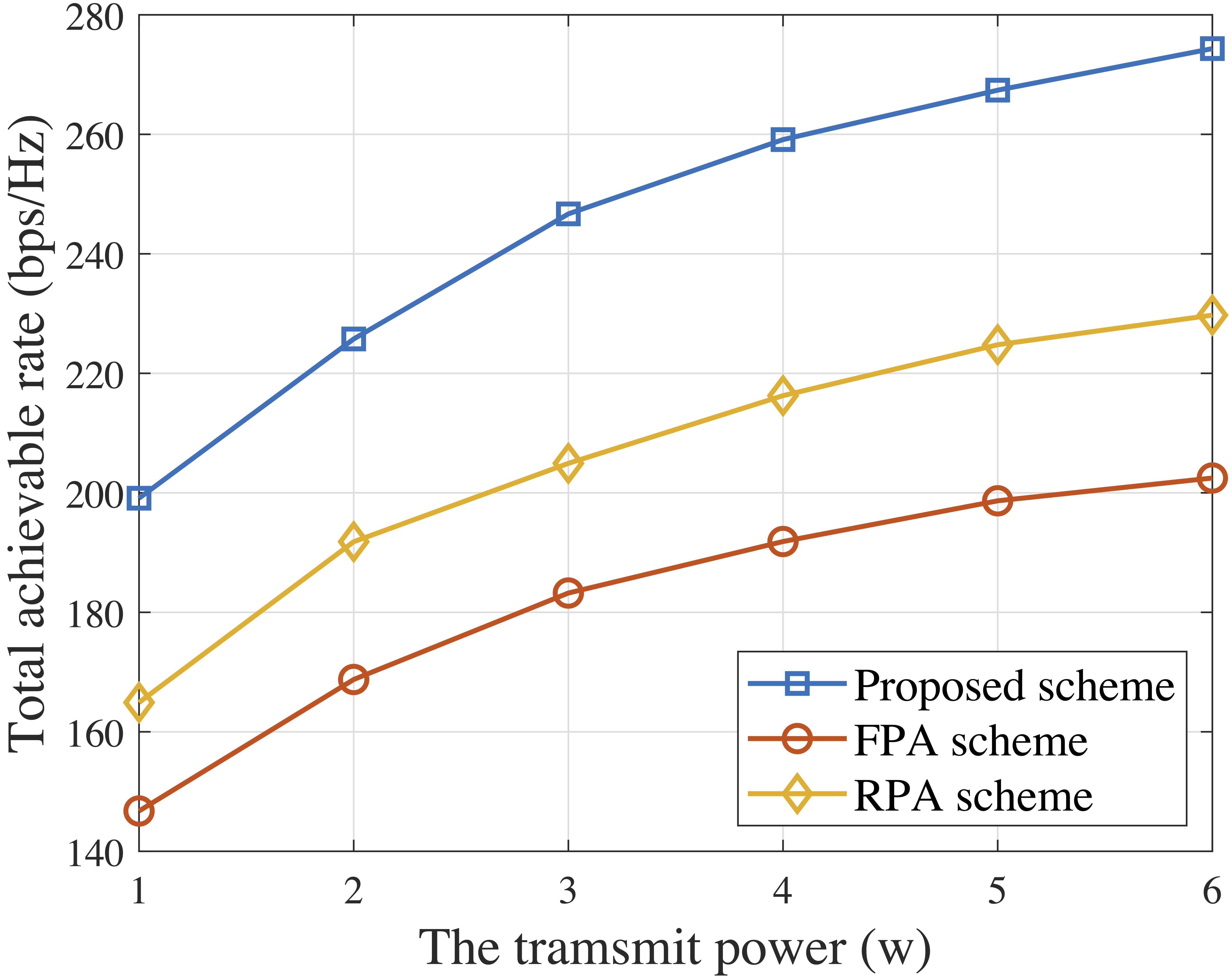}
	\caption{The transmit power versus the achievable data rate.}
	\label{fig:transmit_power}
\end{figure}
The transmit power of the ULAP versus the total achievable data rate is illustrated in Fig.~\ref{fig:transmit_power}.
As we can observe, the proposed scheme always achieves the highest data rate compared with the two benchmarks due to the efficient optimization of joint information and sensing beamforming and antenna positions.
This also verifies that the beamforming gain benefited by the MA array optimization. The data rate of the RPA scheme is higher than that of the FPA scheme. The reason is that, although the positions of antennas are not optimized, the beamforming gain is also obtained due to the spatial degree of freedom of the antenna.

\begin{figure}[!htbp]
	\centering
	\includegraphics[width=0.578\linewidth]{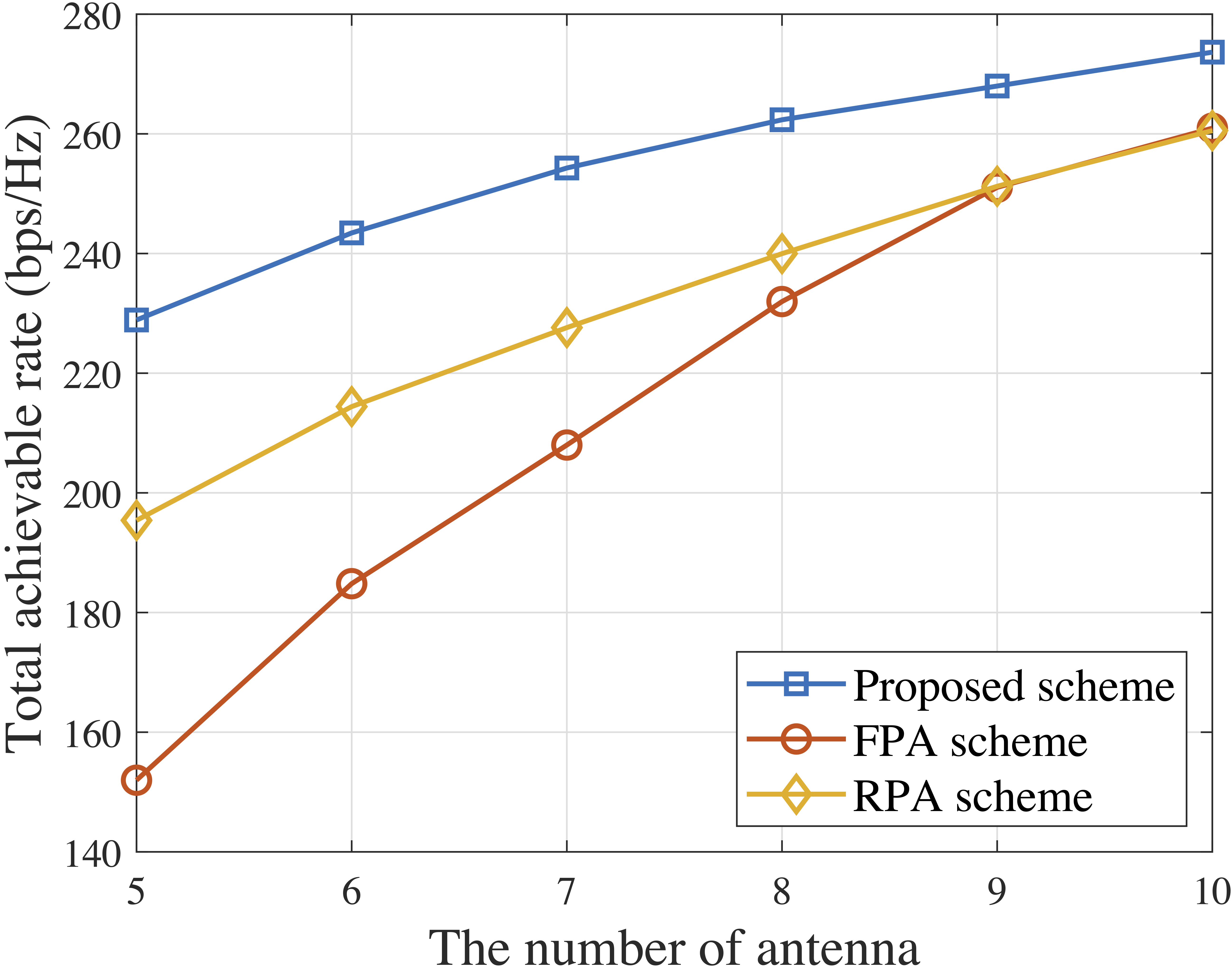}
	\caption{The number of antenna versus the achievable data rate.}
	\label{fig:antenna_number}
\end{figure}
As illustrated in Fig.~\ref{fig:antenna_number}, the number of antennas versus the total achievable data rate is presented.
We can observe that our proposed scheme can always obtain the highest data rate performance, which
confirms the robustness of the proposed scheme when the number of antenna increases. It is worth noting that, the FPA scheme achieves similar data rate gain when the number of the antennas is larger than $9$ compared with the RPA scheme. The reason is that, the interference among users increases as the number of antennas increases. Meanwhile, the FPA scheme has a better performance trade-off compared with the RPA scheme,
while the spatial beamforming gain by the randomly-moved antennas is negated by the increased interference among users.

\begin{figure}[!htbp]
	\centering
	\includegraphics[width=0.578\linewidth]{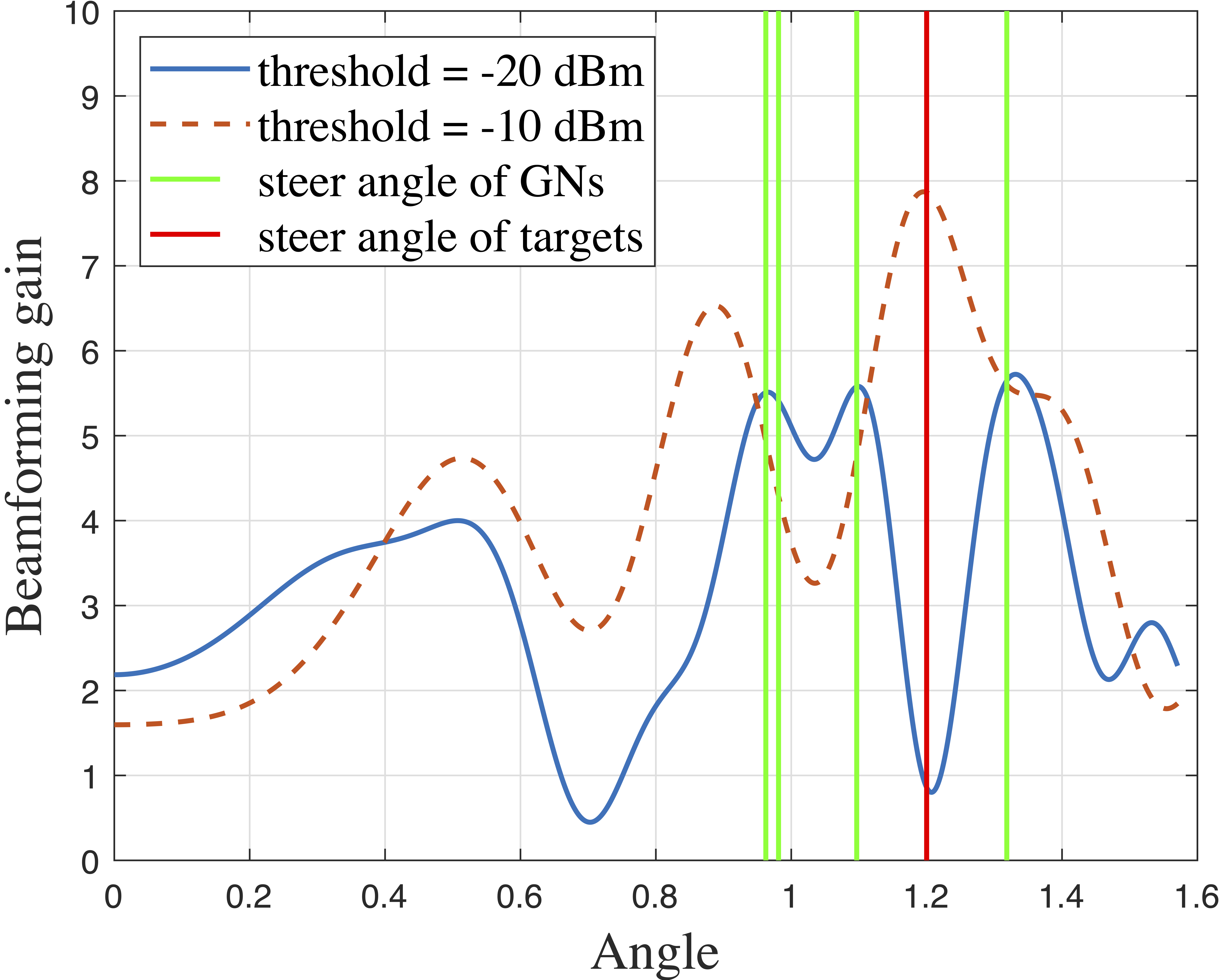}
	\caption{The beamforming gain versus the angle $\theta \in [0, \frac{\pi}{2}]$.}
	\label{fig:beamgain1}
\end{figure}
The beamforming gain under two different sensing thresholds is presented in Fig.~\ref{fig:beamgain1}.
When the threshold of sensing beampattern gain is small ($-20\mathrm{dBm}$), more communication resources are allocated to GNs for information transmission.
Besides, when the threshold is $-10\mathrm{dBm}$, the sensing beamforming also increases to meet the requirement of the sensing beampattern gain.
Meanwhile, all GNs can receive the optimized beamformings with a high performance gain, which results in the maximization of total data rate. Therefore, the robustness of the proposed AO-based scheme is verified.

\section{Conclusion}
This paper investigates the MA array empowered ISAC over ULAP system for supporting the LAE applications, where the achievable data rate is maximized by jointly optimizing the transmit information and sensing beamforming, and the antenna positions of the MA array, while the requirement of the sensing beampattern gain is satisfied.
Numerical results verify the noticeable performance gain in terms of the achievable data rate and the beamforming gain, compared with two benchmarks.

This paper considers the fixed position of the ULAP, where the maneuver control of the ULAP can be further investigated in future works to fully exploit the mobility of the UAV for the LAE applications.
Besides, lower-complexity methods can be further studied in the ULAP scenarios for real-time control and optimization.

\bibliographystyle{IEEEtran}
\bibliography{lap_main}

\begin{thebibliography}{10}
\providecommand{\url}[1]{#1}
\csname url@samestyle\endcsname
\providecommand{\newblock}{\relax}
\providecommand{\bibinfo}[2]{#2}
\providecommand{\BIBentrySTDinterwordspacing}{\spaceskip=0pt\relax}
\providecommand{\BIBentryALTinterwordstretchfactor}{4}
\providecommand{\BIBentryALTinterwordspacing}{\spaceskip=\fontdimen2\font plus
\BIBentryALTinterwordstretchfactor\fontdimen3\font minus \fontdimen4\font\relax}
\providecommand{\BIBforeignlanguage}[2]{{%
\expandafter\ifx\csname l@#1\endcsname\relax
\typeout{** WARNING: IEEEtran.bst: No hyphenation pattern has been}%
\typeout{** loaded for the language `#1'. Using the pattern for}%
\typeout{** the default language instead.}%
\else
\language=\csname l@#1\endcsname
\fi
#2}}
\providecommand{\BIBdecl}{\relax}
\BIBdecl

\bibitem{cheng2024networked}
G.~Cheng, X.~Song, Z.~Lyu, and J.~Xu, ``Networked {ISAC} for low-altitude economy: Transmit beamforming and {UAV} trajectory design,'' \emph{arXiv:2405.07568}, 2024.

\bibitem{8918497}
Y.~Zeng, Q.~Wu, and R.~Zhang, ``Accessing from the sky: A tutorial on {UAV} communications for {5G} and beyond,'' \emph{Proc. IEEE}, vol. 107, no.~12, Dec. 2019.

\bibitem{9737357}
F.~Liu, Y.~Cui, C.~Masouros, J.~Xu, T.~X. Han, Y.~C. Eldar, and S.~Buzzi, ``Integrated sensing and communications: Toward dual-functional wireless networks for {6G} and beyond,'' \emph{IEEE J. Sel. Areas Commun.}, vol.~40, no.~6, pp. 1728--1767, Jun. 2022.

\bibitem{9606831}
Y.~Cui, F.~Liu, X.~Jing, and J.~Mu, ``Integrating sensing and communications for ubiquitous {IoT}: Applications, trends, and challenges,'' \emph{IEEE Network}, vol.~35, no.~5, pp. 158--167, Sep./Oct. 2021.

\bibitem{9456851}
Q.~Wu, J.~Xu, Y.~Zeng, D.~W.~K. Ng, N.~Al-Dhahir, R.~Schober, and A.~L. Swindlehurst, ``A comprehensive overview on {5G}-and-beyond networks with {UAVs}: From communications to sensing and intelligence,'' \emph{IEEE J. Sel. Areas Commun.}, vol.~39, no.~10, pp. 2912--2945, Oct. 2021.

\bibitem{10286328}
L.~Zhu, W.~Ma, and R.~Zhang, ``Movable antennas for wireless communication: Opportunities and challenges,'' \emph{IEEE Commun. Mag.}, pp. 1--7, to appear in 2024.

\bibitem{10278220}
{L. Zhu, W. Ma, and R. Zhang}, ``Movable-antenna array enhanced beamforming: Achieving full array gain with null steering,'' \emph{IEEE Commun. Lett.}, vol.~27, no.~12, pp. 3340--3344, Dec. 2023.

\bibitem{8663615}
Y.~Zeng, J.~Xu, and R.~Zhang, ``Energy minimization for wireless communication with rotary-wing {UAV},'' \emph{IEEE Trans. Wireless Commun.}, vol.~18, no.~4, pp. 2329--2345, Apr. 2019.

\bibitem{8489918}
L.~Xie, J.~Xu, and R.~Zhang, ``Throughput maximization for {UAV}-enabled wireless powered communication networks,'' \emph{IEEE Internet Things J.}, vol.~6, no.~2, pp. 1690--1703, Apr. 2019.

\bibitem{9604506}
X.~Zhang, W.~Luo, Y.~Shen, and S.~Wang, ``Average {AoI} minimization in {UAV}-assisted {IoT} backscatter communication systems with updated information,'' in \emph{Proc. IEEE Ubiquitous Intell. Comput., (UIC)}, 2021, pp. 123--130.

\bibitem{9273074}
W.~Luo, Y.~Shen, B.~Yang, S.~Wang, and X.~Guan, ``Joint {3-D} trajectory and resource optimization in multi-{UAV}-enabled {IoT} networks with wireless power transfer,'' \emph{IEEE Internet Things J.}, vol.~8, no.~10, pp. 7833--7848, May 2021.

\bibitem{10380513}
G.~Cheng, Y.~Fang, J.~Xu, and D.~W.~K. Ng, ``Optimal coordinated transmit beamforming for networked integrated sensing and communications,'' \emph{IEEE Trans. Wireless Commun.}, to appear in 2024.

\bibitem{10462908}
X.~Chen, Z.~Feng, J.~A. Zhang, Z.~Wei, X.~Yuan, P.~Zhang, and J.~Peng, ``Downlink and uplink cooperative joint communication and sensing,'' \emph{IEEE Trans. Veh. Technol.}, to appear in 2024.

\bibitem{10233771}
W.~Liu, Z.~Jin, X.~Zhang, W.~Zang, S.~Wang, and Y.~Shen, ``{AoI}-aware {UAV}-enabled marine {MEC} networks with integrated sensing, computation, and communication,'' in \emph{Proc IEEE/CIC Int. Conf. Commun. China Workshops, (ICCC Workshops)}, 2023, pp. 1--6.

\bibitem{10382465}
Y.~Chen, H.~Hua, J.~Xu, and D.~W.~K. Ng, ``{ISAC} meets {SWIPT}: Multi-functional wireless systems integrating sensing, communication, and powering,'' \emph{IEEE Trans. Wireless Commun.}, to appear in 2024.

\bibitem{9916163}
Z.~Lyu, G.~Zhu, and J.~Xu, ``Joint maneuver and beamforming design for {UAV}-enabled integrated sensing and communication,'' \emph{IEEE Trans. Wireless Commun.}, vol.~22, no.~4, pp. 2424--2440, Apr. 2023.

\bibitem{10382559}
W.~Ma, L.~Zhu, and R.~Zhang, ``Multi-beam forming with movable-antenna array,'' \emph{IEEE Commun. Lett.}, vol.~28, no.~3, pp. 697--701, Mar. 2024.

\bibitem{10504625}
B.~Feng, Y.~Wu, X.-G. Xia, and C.~Xiao, ``Weighted sum-rate maximization for movable antenna-enhanced wireless networks,'' \emph{IEEE Wireless Commun. Lett.}, to appear in 2024.

\bibitem{liu2024}
W.~Liu, X.~Zhang, H.~Xing, J.~Ren, Y.~Shen, and S.~Cui, ``{UAV}-enabled wireless networks with movable-antenna array: Flexible beamforming and trajectory design,'' \emph{arXiv:2405.20746}, 2024.

\bibitem{9124713}
X.~Liu, T.~Huang, N.~Shlezinger, Y.~Liu, J.~Zhou, and Y.~C. Eldar, ``Joint transmit beamforming for multiuser {MIMO} communications and {MIMO} radar,'' \emph{IEEE Trans. Signal Process.}, vol.~68, pp. 3929--3944, 2020.

\bibitem{10086626}
H.~Hua, J.~Xu, and T.~X. Han, ``Optimal transmit beamforming for integrated sensing and communication,'' \emph{IEEE Trans. Veh. Technol.}, vol.~72, no.~8, pp. 10\,588--10\,603, Aug. 2023.

\bibitem{10508198}
J.-M. Kang, ``Deep learning enabled multicast beamforming with movable antenna array,'' \emph{IEEE Wireless Commun. Lett.}, to appear in 2024.

\end{thebibliography}
\end{document}